\documentclass[11pt, a4paper,superscriptaddress,aps,nofootinbib,  ]{revtex4}
\usepackage{amsmath,amssymb,graphicx}
\makeatletter
\usepackage[active]{srcltx}
\usepackage{graphicx,color}
\usepackage{changebar}
\usepackage{hyperref}
\hypersetup{
	colorlinks=true,
	urlcolor= blue,
	citecolor=blue,
	linkcolor= blue,
	bookmarks=true,
	bookmarksopen=false,
}
\usepackage[T1]{fontenc}
\usepackage{ae}
\usepackage[ansinew]{inputenc}
\usepackage{amsmath}
\usepackage{braket}
\usepackage{mathtools}
\usepackage{slashed}
\usepackage{empheq}
\usepackage{tikz}
\usepackage{multirow}
\usetikzlibrary{decorations.pathmorphing}
\usepackage[caption = false]{subfig}
 \usetikzlibrary{quotes,angles}

\usepackage{graphicx}
\usepackage{amsfonts}
\usepackage{amsmath}
\usetikzlibrary{arrows} 
\usetikzlibrary{decorations.markings}

\begin{document}
\title{ Gravitational lensing in black-bounce spacetimes}

\author{J. R. Nascimento}
\email[]{jroberto@fisica.ufpb.br}
\affiliation{Departamento de F\'{\i}sica, Universidade Federal da 
	Para\'{\i}ba,\\
	Caixa Postal 5008, 58051-970, Jo\~ao Pessoa, Para\'{\i}ba, Brazil}

\author{A. Yu. Petrov}
\email[]{petrov@fisica.ufpb.br}
\affiliation{Departamento de F\'{\i}sica, Universidade Federal da 
	Para\'{\i}ba,\\
	Caixa Postal 5008, 58051-970, Jo\~ao Pessoa, Para\'{\i}ba, Brazil}

\author{P. J. Porf\'{i}rio}\email[]{pporfirio89@gmail.com}
\affiliation{Departamento de F\'{\i}sica, Universidade Federal da 
	Para\'{\i}ba,\\
	Caixa Postal 5008, 58051-970, Jo\~ao Pessoa, Para\'{\i}ba, Brazil}

\author{A. R. Soares}
\email[]{adriano2da@gmail.com}
\affiliation{Departamento de F\'{\i}sica, Universidade Federal da 
	Para\'{\i}ba,\\
	Caixa Postal 5008, 58051-970, Jo\~ao Pessoa, Para\'{\i}ba, Brazil}

\begin{abstract}
In this work, we calculate the deflection angle of light in a spacetime that interpolates between regular black holes and traversable wormholes, depending on the free parameter of the metric. Afterwards, this angular deflection is substituted into the lens equations which allows to obtain physically measurable results, such as the position of the relativistic images and the magnifications. 

\end{abstract}

\maketitle
 
 \section{INTRODUCTION}
 \label{Int.}

 The angular deflection of light when passing through a gravitational field was one of the first predictions of the General Relativity (GR). Its confirmation played a role of a milestone for GR being one of the most important tests for it \cite{Eddington,Crispino2019}. Then, gravitational lenses have become an important research tool in astrophysics and cosmology \cite{Einstein1936, Liebes}, allowing studies of the distribution of structures \cite{Mellier1999, Schneider2001}, dark matter \cite{Kaiser1993} and some other topics \cite{HSV, Schmidt2008, Guzik2010, Sakalli2017, Jusufi2017, KJusufi2017, Goulart2018, Ono2018}. Like as in the works cited earlier, the prediction made by Einstein was developed in the weak field approximation, that is, when the light ray passes at very large distance from the source {which generates the gravitational lens. Under the phenomenological point of view, the recent discovery of gravitational waves by the LIGO-Virgo collaboration \cite{LIGO1, LIGO2, LIGO3} opened up a new route of research, that is, to explore new cosmological observations by probing the Universe with gravitational waves, in particular, studying effects of gravitational lensing in the weak field approximation (see e.g. \cite{Contigiani:2020yyc, Mukherjee:2019wcg} and references therein).

On the other hand, in the strong field regime, the light rays pass very close to the source of the lens, usually given by a compact object like a black hole; in this case, the angular deflection of the light becomes very large. The first studies addressing gravitational lenses in the strong field regime came up with the  paper \cite{Darwin} in Schwarszchild spacetime and, later, for general spherically symmetric and static spacetimes \cite{Atkinson1965}. The studies along this line have been intensified in the last years, in particular, driven by the experimental observations that indicate a supermassive black hole living at the center of the Milky Way \cite{Eisenhauer:2005cv}. More recently, their direct observations have been made by the Event Horizon Telescope (EHT) collaboration which published pictures of the black hole shadow at the center galaxy M87 \cite{Akiyama:2019cqa, Akiyama:2019brx, Akiyama:2019sww,Akiyama:2019bqs,Akiyama:2019fyp,Akiyama:2019eap}. Accordingly, the gravitational field surrounding these objects is intense and, therefore, the weak field approximation fails to apply. In this case a strong field approach must be taken into account.
 In the strong field limit the mathematical treatment of gravitational lenses becomes cumbersome for tackling. However, in recent years significant efforts have been made toward a full analytical treatment. In this limit, Virbhadra and Ellis got a simple lens equation for a galactic supermassive black hole in an asymptotically flat background which can be applied properly to large deflections of light \cite{Virbhadra-Ellis-2000, Perlick2008}. They found that, apart from primary and secondary images, there exist an infinite number of images on both sides of the optic axis, as a consequence of the strong gravitational field experienced by the light rays. Fritelli \textit{et al.} \cite{Frittelli:1999yf}, without mentioning any sort of background, set up an exact lens equation for the Schwarszchild spacetime. Bozza \textit{et al.} \cite{Boz-Cap2001} obtained a background-independent}   analytical expression for the angle of light deflection in the strong field limit, that is, very close to the photon sphere and then applied for the Schwarszchild spacetime \cite{Bozza2002}. Posteriorly, Tsukamoto improved the result found by Bozza \cite{Tsukamoto2017}. The angular deflection of light in the strong field limit has been studied in several contexts, including the Reissner-Nordstr\"{o}m spacetime \cite{Eiroa2002}, rotating solutions \cite{rotacao}, wormholes \cite{WH}, topological defects \cite{DfTp}, modified theories of gravity \cite{TM}, regular black holes \cite{Reg}, naked singularities \cite{naked} and braneworld scenarios \cite{extra}.
 
 It is well known that some classical solutions in GR suffer from physical singularities in the strong field limit (at the Planck scale) which means that GR breaks down. On the other hand, GR also fails at large scales since it cannot explain the accelerating expansion of the Universe \cite{LCDM}. In the absence of a full theory of quantum gravity, alternative theories of gravity have been explored as effective theories. In particular, we call attention to the Einstein-Born-Infeld (EiBI) gravity in which many studies demonstrated presence of solutions free from singularities and without exotic matter \cite{conj,PhyRep2018,Soares2019, Soares2020,GR2012,Shaikh2015,Shaikh2018}, for example, traversable wormholes and regular black holes. In the same spirit, Simpson and Visser introduced the so-called  black-bounce traversable wormhole spacetime \cite{SimVisser2019}, hereinafter referred to as the black-bounce spacetime, where, depending on the relationship between model parameters, the metric can generate solutions for regular black holes and traversable wormholes, that is, this metric is free from the singularity problem. The authors calculated several quantities related to the aforementioned spacetime and, in the GR context, they showed that the solution is supported by exotic matter and then violating the energy conditions. Generalizations of \cite{SimVisser2019} have been put forward in the literature, for example, in \cite{SimpMViss}, the authors considered a non-static evolving version and, in \cite{thinshell}, the authors considered a thin-shell setup for black-bounce spacetimes, in \cite{Churi}, the authors examined the quasinormal modes of black-bounce spacetimes.

 In this paper, we pursue two goals. First, we must calculate the angular deflection of light, both in the weak field limit and in the strong field limit, in the black-bounce spacetime introduced by Simpson and Visser. Second, with the angular deflection of light, we will investigate the observables related to the respective gravitational lensing.
 
 The paper is organized as follows: in the section \ref{SFD}, we will present the metric that describes the black-bounce space-time and use the methodology developed by Bozza \cite{Bozza2002} and Tsukamoto \cite{Tsukamoto2017} to obtain the expression for the angular deflection of light in the strong field limit. In the section \ref{LEO}, we introduce an expression for a deflection of light in the strong field limit in the lens equation, and afterwards, we study the observables related to corresponding relativistic images. 
 Finally, in the section \ref{conc} we discuss our main results and present our conclusions. Throughout this paper we adopt geometrized units, $G=c=1$.
  
 \section{ BLACK-BOUNCE SPACETIMES AND DEFLECTION OF LIGHT}\label{SFD}
 In this section, we start with discussing the metric introduced by Simpson and Visser in \cite{SimVisser2019}. Next, we will use the results obtained in \cite{Bozza2002, Tsukamoto2017} to find the expression for the angular deflection of light in the strong field limit, i.e., when the light ray passes very closely to the photon sphere.   
 
  The line element of the metric describing the  black-bounce spacetime  is given by
  \begin{equation}\label{bm}
  ds^2=-\left(1-\frac{2M}{\sqrt{x^2+a^2}}\right)dt^2+\left(1-\frac{2M}{\sqrt{x^2+a^2}}\right)^{-1}dx^2+(x^2+a^2)\left(d\theta^2+\sin^2\theta d\phi^2\right) \ ,
  \end{equation}
  where $a$ is a parameter characterizing the bounce length scale. We note that this is a one-parametric modification of the standard Schwarzschild solution, so when we take $a=0$, (\ref{bm}) reduces to the  Schwarzschild solution. Furthermore, it is a static and spherically symmetric space-time. The domain of radial and temporal coordinates is: $-\infty<x< \infty$, $-\infty<t<\infty$. It is worth noting that in the case $a\neq 0$, the metric is finite throughout and, as a consequence, it describes non-trivial topological structures as regular black holes and wormholes depending on the parameter $a$. For example, adjusting the bounce parameter to assume values in the range, $\frac{a}{2M}>1$, the metric describes a traversable wormhole and then particles are allowed to go from the one its side to the other one. The other possibilities are: $\frac{a}{2M}=1$, which describes a one-way (non-traversable) wormhole, in the sense that particles cannot go through, to the other side since there exists a horizon at $r=0$, and $\frac{a}{2M}<1$, when the geometry corresponds to a regular black hole with horizons located at $r_{\pm}=\sqrt{2M^2-a^2}$ (for more details on this geometry, see \cite{SimVisser2019,SimpMViss}). Of course, this geometry is only supported by exotic matter sources in GR, thus violating the energy conditions. However, such structures have been found in alternative theories of gravity with no the requirement of exotic matter, see for example \cite{Olmo:2013gqa}; thereby this, by itself, is a well-motivated argument to explore the black-bounce metrics. In the following, we do a brief discussion of the weak field approach and then we concentrate our efforts on the influence of the black-bounce parameter $a$ in the deflection of light in the strong field regime. 
	
Now, to study the deflection of light let us consider the geodesics of particles moving in the black-bounce background. To proceed further, we shall consider the equation for null geodesics (for our purposes it suffices to consider null geodesics) with a related affine parameter $\lambda$ in the equatorial plane $ \theta=\frac{\pi}{2}$ of the spacetime (\ref{bm}) is given by
  \begin{equation}\label{Eg}
  	-\left(1-\frac{2M}{\sqrt{x^2+a^2}}\right)\left(\frac{dt}{d\lambda}\right)^2+\left(1-\frac{2M}{\sqrt{x^2+a^2}}\right)^{-1}\left(\frac{dx}{d\lambda}\right)^2+\left(x^2+a^2\right)\left(\frac{d\phi}{d\lambda}\right)^2=0 \ .
  \end{equation}
  It can still be shown that the following quantities are conserved
\begin{equation}\label{E}
	E=\left(1-\frac{2M}{\sqrt{x^2+a^2}}\right)\left(\frac{dt}{d\lambda}\right)
\end{equation} 
and
\begin{equation}\label{L}
	L=\left(x^2+a^2\right)\frac{d\phi}{d\lambda} \ ,
\end{equation}
where $E$ is the energy and $L$ is the angular momentum. Substituting (\ref{L}) and (\ref{E}) in (\ref{Eg}), we can then show that the equation for radial geodesics is given by
\begin{equation}\label{gr}
	\left(\frac{dx}{d\lambda}\right)^2=E^2-\frac{L^2}{(x^2+a^2)}\left(1-\frac{2M}{\sqrt{x^2+a^2}}\right) \ .
\end{equation}
It is known that the equation (\ref{gr}) describes a particle with energy $E$ in a one-dimensional path governed by the potential $V_{eff}(x)=\frac{L^2}{(x^2+a^2)}\left(1-\frac{2M}{\sqrt{x^2+a^2}}\right)$ \cite{Carrolbook}. Therefore, circular orbits (photon sphere) happen for extremum of $V_{eff}(x)$, i.e., where $\frac{dV_{eff}(x)}{dx}=0$. Thus, the photon sphere radius $x_m$ is given by
\begin{equation}\label{rps}
x_m=\pm\sqrt{(3M)^2-a^2} \ .
\end{equation}
The signs $+$ and $-$ correspond to each side of the solution. As the solution is symmetric with respect to the throat $x=0$, we can choose any side to calculate the light deflection, and the result is the same. So, without loss of generality, let us focus on the side where $x_m>0$.  In the black hole case, where $\frac{a}{2M}<1$, there will always be a photon sphere. In the wormhole case, there is only a photon sphere if $a<3M$. In any case, we observed that the radius of the photon sphere is smaller than in the Schwarzschild solution, where $a=0$.


We admit that the photon starts its journey from an asymptotically flat region, approaches the object (black hole or wormhole) and then deviates at the closest distance $x=x_0$, with $x_0>x_m$, and goes to another asymptotically flat region of spacetime. In $x_0$ we should have $V_{eff}=E^2$, then the (\ref{gr}) implies
\begin{eqnarray}\label{b}
	\frac{E^2}{L^2}&=&\frac{1}{(x_0^2+a^2)}\left(1-\frac{2M}{\sqrt{x_0^2+a^2}}\right) \ , \nonumber\\
	b(x_0)&=&\left[\frac{1}{(x_0^2+a^2)}\left(1-\frac{2M}{\sqrt{x_0^2+a^2}}\right)\right]^{-1/2} \ ,
\end{eqnarray}
where we define the critical impact parameter $b\equiv\frac{L}{E}$. Replacing (\ref{L}) in (\ref{gr}), we have
\begin{equation}
	\frac{d\phi}{dx}=\left[\frac{(x^2+a^2)^2}{b^2}-(x^2+a^2)\left(1-\frac{2M}{\sqrt{x^2+a^2}}\right)\right]^{-1/2} \ .
\end{equation}
The angular deflection of light $\alpha(x_0)$ then will be given by 
\begin{equation}\label{defa}
	\alpha(x_0)=2\int_{x_0}^{\infty}\left[\frac{(x^2+a^2)^2}{b^2}-(x^2+a^2)\left(1-\frac{2M}{\sqrt{x^2+a^2}}\right)\right]^{-1/2} \ dx-\pi \ .
\end{equation}
Let us calculate the angular deflection (\ref{defa}) initially in the weak field limit, 
we follow the procedure analogous to \cite{Wald}. For this, we introduce the change of variable: $u=\frac{1}{\sqrt{x^2+a^2}}$, so we are left with
\begin{eqnarray}\label{alpha}
	\alpha&=&2\int_{0}^{u_0}\left[(1-a^2u^2)\left(\frac{1}{b^2}-u^2(1-2Mu)\right)\right]^{-1/2}du-\pi \nonumber\\
	\alpha&=&2\int_{0}^{u_0}\left[(1-a^2u^2)\left(u_0^2(1-2Mu_0)-u^2(1-2Mu)\right)\right]^{-1/2}du-\pi \ .
\end{eqnarray}
When moving from the first to the second line, we use (\ref{b}) to express $b$ in terms of $u_0$. Taking into account the terms up to the first order in $M$ and second order in $a$ and then performing the integration over $u$ in (\ref{alpha}), we find
 \begin{equation}
 	\alpha(b)\simeq\frac{4M}{b}+\frac{a^2\pi}{4b^2}+\frac{a^2M(16-3\pi)}{6b^3}+\mathcal{O}(Ma^3) \ .
 \end{equation}
 The first term is the angular deflection in the standard Schwarzschild spacetime, the second term comes from of the bounce parameter and so on.
Therefore, being compared to the Schwarzschild solution, the angular deflection of light in the weak field approximation will be greater due to the quadratic contribution of the bounce parameter $a$. However, this increase only occurs in the second order in $a$, therefore, we expect that for practical purposes, the observables, such as the  images positions and magnifications, will remain the same as in the Schwarzschild solution.

\subsection{EXPANSION FOR DEFLECTION OF LIGHT IN THE STRONG FIELD LIMIT}
In this subsection we will consider the angular deflection of light in the strong field limit.
As the ray of light approaches the photon sphere, the deflection of the light increases, and at the limit $x_0\to x_m$ the angular deflection diverges logarithmically. The approximate expression for the angular deflection of light in terms of the impact parameter $\alpha(b)$ was obtained in  \cite{Tsukamoto2017} for a static and spherically symmetric spacetime, with the following generic line element:
 \begin{equation}\label{gm}
 ds^2=-A(x)dt^2+B(x)dx^2+C(x)\left(d\theta^2 +\sin^2\theta d\phi^2\right).
 \end{equation}
 Note that in our specific case,
  \begin{equation}\label{AC}
  A(x)=\frac{1}{B(x)}=1-\frac{2M}{\sqrt{x^2+a^2}}\quad\text{and}\quad C(x)=x^2+a^2  \  .
  \end{equation}
From (\ref{gm}), it can be shown that the angular deflection of light $\alpha(x_0)$ is given by
\begin{equation}
\alpha(x_0)=I(x_0)-\pi \ ,
\end{equation}
  where
  \begin{equation}\label{I}
  I(x_0)=2\int_{x_0}^{\infty}\frac{dx}{\sqrt{\frac{R(x)C(x)}{B(x)}}}\quad\text{with} \quad R(x)=\frac{A_0C}{AC_0}-1 \ .
  \end{equation}
  In the above expression, the subscript $``0"$ means that the function is evaluated at the turning point $x=x_0$, for example $A_0=A(x_0)$.
 Realize that we already got the expression for $I(x_0)$ in (\ref{defa}), where
 \begin{equation}
 	\frac{R(x)C(x)}{B(x)}=\frac{(x^2+a^2)^2}{b^2}-(x^2+a^2)\left(1-\frac{2M}{\sqrt{x^2+a^2}}\right) \ .
 \end{equation}
    After introducing the variable 
   \begin{equation}\label{v}
   z=1-\frac{x_0}{x} \ ,
   \end{equation}
   $I(x_0)$ is written as
   \begin{equation}
   I(x_0)=\int_{0}^{1}f(z,x_0)dz \ ,
   \end{equation}
   where 
   \begin{equation}\label{f-G}
   f(z,x_0)=\frac{2x_0}{\sqrt{G(z,x_0)}}	\quad\text{with}\quad G(z, x_0)=\frac{RC}{B}(1-z)^4.
   \end{equation}
   The integral $I(x_0)$  can be split into a divergent part $I_D(x_0)$ and a regular part $I_R(x_0)$. The divergent part is given by
   \begin{equation}
   I_D(x_0)=\int_{0}^{1}f_D(z,x_0) \ dz \ ,
   \end{equation}
   where $f_D(z,x_0)=\frac{2x_0}{\sqrt{c_1z+c_2z^2}}$, with $c_1$ and $c_2$ are coefficients of the series expansion of the function $G(z,x_0)$ up to the second order in $z$ (\ref{f-G}). The regular part is obtained from $I(x_0)$ subtracting the divergent part, i.e.,
   \begin{equation}\label{regp}
   I_R(x_0)=\int_{0}^{1}f_R(z,x_0) \ dz, \quad f_R(z,x_0)=f(z,x_0)-f_D(z,x_0) \ .
   \end{equation}
   The deflection angle of the light in the strong limit $x_0\to x_m$ is given by \cite{Tsukamoto2017,Bozza2002} 
   \begin{equation}\label{angdefle}
   \alpha(b)=-\bar{a}\log\left(\frac{b}{b_c}-1\right)+\bar{b}+\mathcal{O}[(b-b_c)\log(b-b_c)] \ ,
   \end{equation}
    where $b_c=\lim\limits_{x_0\to x_m}b(x_0)$, 
   \begin{equation}\label{abarra}
   \bar{a}=\sqrt{\frac{2B_mA_m}{C''_mA_m-C_mA''_m}} \ , 
   \end{equation}
   and 
   \begin{equation}\label{bbarra}
   \bar{b}=\bar{a}\log\left[x_m^2\left(\frac{C''_m}{C_m}-\frac{A''_m}{A_m}\right)\right]+I_R(x_m)-\pi \ .
   \end{equation}
  The superscript ${\bf X}''_m$ means second derivative of ${\bf X}(x)$ in relation to $x$ evaluated in $x=x_m$, i.e., ${\bf X}''_m=\frac{d^2\mathbf{X}(x)}{dx^2}\Big|_{x=x_m}$.

\begin{figure}[h]
      	\centering
      	\includegraphics[height=6cm]{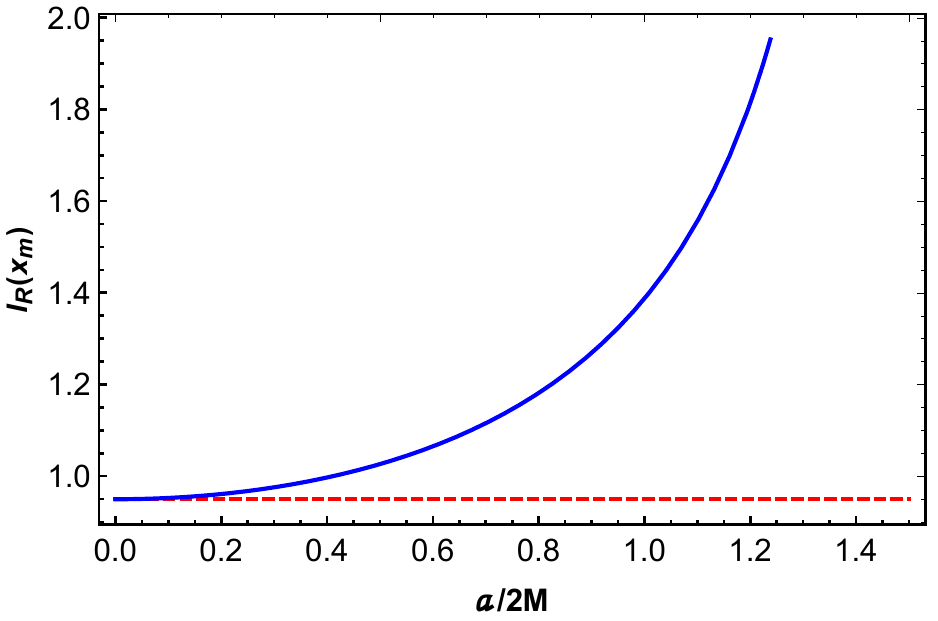}
      	\caption{The solid (blue) curve describes $I_R(x_m)$ in function of $a/(2M)$. The dashed (red) curve is the value of $I_R(x_m)$ in the Schwarzschild solution, when $a=0$, i.e., $\approx 0.949$. } 
      	\label{Ir}
      \end{figure}

\begin{figure}
      	\subfloat[]{\includegraphics[width = 3in]{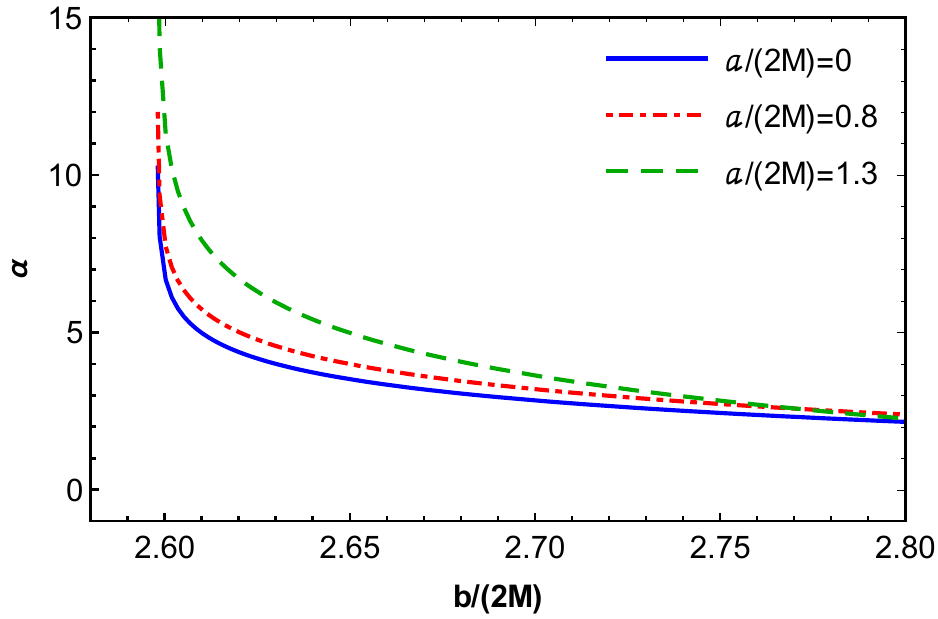}} 
      	\subfloat[]{\includegraphics[width = 3in]{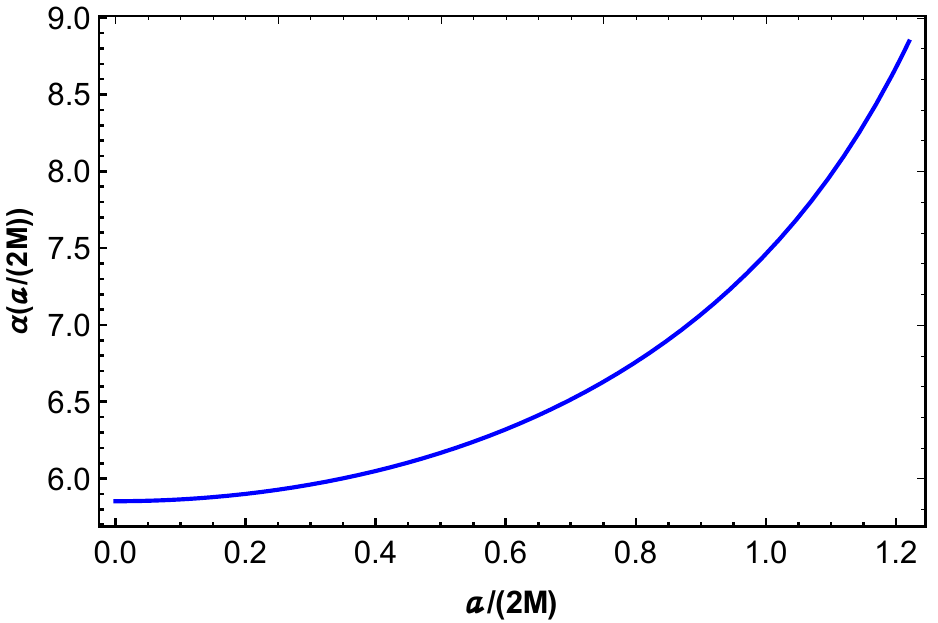}} 
      	\caption{In (a), angular deflection of light as a function of the impact parameter for various values of the bounce parameter $a$. In (b),  angular deflection of light as a function of $\frac{a}{2M}$, for $\frac{b}{2M}=\frac{3\sqrt{3}}{2}+0.005$.  }
      	\label{CEi}
      \end{figure}

  So let us calculate the expansion coefficients (\ref{angdefle}) for the black-bounce spacetime (\ref{AC}), starting with the critical impact parameter. Substituting (\ref{rps}) in (\ref{b}), we have 
  \begin{equation}\label{bc}
  	b_c=3\sqrt{3}M \ ,
  \end{equation}
 as well as in the Schwarzschild solution. The next step is to calculate $\bar{a}$. Substituting (\ref{AC}) in (\ref{abarra}), we have
   \begin{eqnarray}
  \bar{a}(x_m)=\left[\frac{\left(a^2+x_m^2\right)^{3/2}}{a^2 \left(\sqrt{a^2+x_m^2}-3 M\right)+x_m^2 \sqrt{a^2+x_m^2}}\right]^{\frac{1}{2}}.
	\end{eqnarray}
Now using the fact that	$x_m$ is given by (\ref{rps}), we arrive at
	\begin{eqnarray}\label{abarra2}
  \bar{a}&=&\frac{3M}{\sqrt{(3M)^2-a^2}} \ .
  \end{eqnarray}
   Substituting (\ref{AC}) in (\ref{bbarra}) and using (\ref{rps}), we get
   \begin{equation}\label{bbarra2}
   \bar{b}=\frac{3M}{\sqrt{(3M)^2-a^2}}\log\left[\frac{2(a^2-9M^2)^2}{27M^4}\right]     +I_R(x_m)-\pi \ .
   \end{equation}
    We still need to calculate the integral $I_R(x_m)$, given by (\ref{regp}). For this, we first write the function $G(z,x_m)$, equation (\ref{f-G}), which is given by
    \begin{eqnarray}\label{Gzxm}
    	G(z,x_m)&=&2 M (z-1)^4 \sqrt{\frac{a^2 (z-2) z+9 M^2}{(z-1)^2}}\nonumber\\
    	&+&(z-1)^2 \left(a^2-9 M^2\right)+\frac{\left(a^2 (z-2)
    		z+9 M^2\right)^2}{27 M^2}-a^2 (z-1)^4 \ .
    \end{eqnarray}
    Expanding (\ref{Gzxm}) in powers of $z$ up to second order, close to $z=0$, we have
    \begin{equation}\label{Gzxma}
    	G(z,x_m)\simeq\frac{z^2 \left(a^2-9 M^2\right)^2}{9 M^2} \ .
    \end{equation}
    With (\ref{Gzxma}) and (\ref{Gzxm}) we can  then get the regular part,
     \begin{eqnarray}\label{IR}
     I_R(x_m)&=& \int_{0}^{1}\frac{2 \sqrt{9 M^2-a^2} \ dz}{\sqrt{2 M (z-1)^4 \sqrt{\frac{a^2 (z-2) z+9 M^2}{(z-1)^2}}+(z-1)^2 \left(a^2-9 M^2\right)+\frac{\left(a^2 (z-2) z+9 M^2\right)^2}{27 M^2}-a^2 (z-1)^4}}\nonumber\\
     &-&\int_{0}^{1}\frac{6 M \sqrt{9 M^2-a^2}\ dz}{z \sqrt{\left(a^2-9 M^2\right)^2}} \ .
     \end{eqnarray}
     We can evaluate numerically (\ref{IR}). The result is plotted in Fig.\ref{Ir}. As we can see $I_R(x_m)$ increases with the value of $a$.
      Therefore, replacing (\ref{bc}), (\ref{abarra}) and (\ref{bbarra}) in (\ref{angdefle}), we have the deflection angle in strong field limit,
      \begin{eqnarray}\label{angular}
      \alpha(b)&=&-\frac{3M}{\sqrt{(3M)^2-a^2}}\log\left(\frac{b}{3\sqrt{3}M}-1\right)+\frac{3M}{\sqrt{(3M)^2-a^2}}\log\left[\frac{2(a^2-9M^2)^2}{27M^4}\right]\nonumber\\     &+&I_R(x_m)-\pi+\mathcal{O}\left[(b-b_c)\log(b-b_c)\right] \ .
      \end{eqnarray}
      In the Fig.\ref{CEi} (a), we presented the angular deflection of light (\ref{angular}) in terms of the function of $\frac{b}{2M}$ for several values of $\frac{a}{2M}$. In the Fig.\ref{CEi} (b), the  $ \alpha(b)$ in $\frac{b}{2M}=\frac{3\sqrt{3}}{2}+0.005$ is given as a function of $\frac{a}{2M}$.  As we can see from the Fig. \ref{CEi}, the angular deflection increases with the bounce parameter value $a$. When we take $a=0$ in (\ref{angular}) we recover the angular deflection of light obtained in Schwarzschild spacetime \cite{Bozza2002}, as it should be, i.e.,
      \begin{equation}
      	\alpha(b)=-\log\left(\frac{b}{3\sqrt{3}M}-1\right)+\log(6)+0.9496-\pi+\mathcal{O}\left[(b-b_c)\log(b-b_c)\right] \ .
      \end{equation}

       \section{Lens equation and observables} \label{LEO}
       
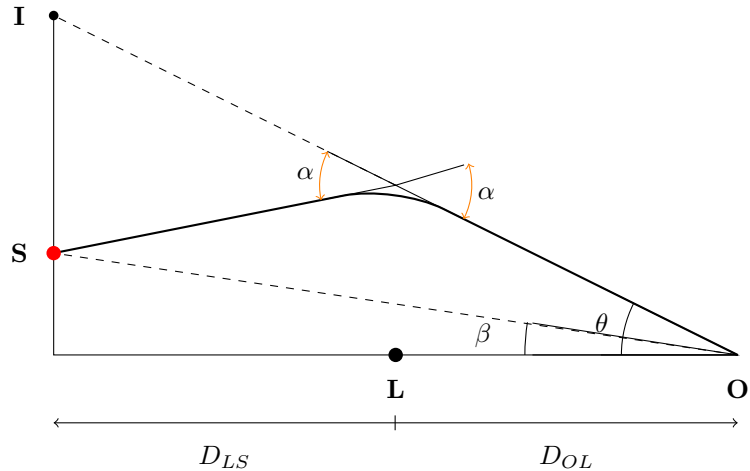
\begin{figure}
       	\centering
       	\begin{tikzpicture}[scale=0.9]
       	\node (I)    at ( 5,-0.5)   {{\bf L}};
       	\node (II)    at ( 10,-0.5)   {{\bf O}};
       	\node (II)    at ( -0.5,1.5)   {{\bf S}};
       	\node (II)    at ( -0.5,5)   {{\bf I}};
       	\node    at ( 2.5,-1.5)   {$D_{LS}$};
       	\node    at ( 7.5,-1.5)   {$D_{OL}$};
       	
       	\draw (10,0)--(0,0)--(0,5);
       	\draw [thick,rounded corners=20pt] (0,1.5)--(5,2.5)--(10,0);
       	\draw [dashed] (5,2.5)--(0,5);
       	\draw [dashed](10,0)--(0,1.5);
       	\fill[black] (5,0) circle (3pt);
       	\fill[red] (0,1.5) circle (3pt);
       	\fill[black] (0,5) circle (2pt);
       	\draw
       	(6,2) coordinate (a) 
       	-- (5,2.5) coordinate (b) 
       	-- (6,2.8) coordinate (c) 
       	pic["$\alpha$", draw=orange, <->, angle eccentricity=1.2, angle radius=1cm]
       	{angle=a--b--c};
       	
       	\draw
       	(4,2.3) coordinate (d) 
       	-- (5,2.5) coordinate (e) 
       	-- (4,3) coordinate (f) 
       	pic["$\alpha$", draw=orange, <->, angle eccentricity=1.2, angle radius=1cm]
       	{angle=f--e--d};
       	
       	\draw
       	(8,0) coordinate (g) 
       	-- (10,0) coordinate (h) 
       	-- (8,1) coordinate (i) 
       	pic["$\theta$", draw=black, angle eccentricity=1.2, angle radius=1.53cm]
       	{angle=i--h--g};
       	
       	\draw
       	(7,0) coordinate (g1) 
       	-- (10,0) coordinate (h1) 
       	-- (7,0.47) coordinate (i1) 
       	pic["$\beta$", draw=black, angle eccentricity=1.2, angle radius=2.8cm]
       	{angle=i1--h1--g1};	
       	
       	\draw[ <->,
       	decoration={markings,
       		mark= at position 0.5 with {\arrow{|}},
       	},
       	postaction={decorate}
       	]
       	(0,-1) node[anchor=west] {} -- (10,-1);
       	
       	\end{tikzpicture}
       	\caption{In this Lens diagram, $D_{OL}$ is the distance between the lens $L$ and the observer $O$, and $D_{LS}$ is the distance between the source projection in relation to optical axis and the lens. }
       	\label{LG}
       \end{figure}

       In this section we will derive several quantities related to the deflection of light  in  the strong field limit by the black-bounce spacetime. First, let us define the geometric configuration of the lens shown in the Fig. \ref{LG}. The light emitted by the source $S$ is deflected towards the observer $O$ by the compact object located in $L$. The angular positions of the source and the image seen by the observer are, respectively, $\beta$ and $\theta$, and the angular deflection of light is given by $\alpha$. 
       
       We adopt here the same configuration used in \cite{Boz-Cap2001, Virbhadra-Ellis-2000}, that is, we assume that the source ($S$) is almost perfectly aligned with the lens ($L$), which is precisely the case where relativistic images are more expressive. In this case, the lens equation relating $\theta$ and $\beta$ is given by
       \begin{equation}\label{EqLente}
       \beta=\theta-\frac{D_{LS}}{D_{OS}}\Delta\alpha_{n}\ ,
       \end{equation}
       where $\Delta\alpha_n$ is the deflection angle less all the loops made by the photons before reaching the observer, i.e., $\Delta\alpha_{n}=\alpha-2n\pi$. In this approach, we use the following approximation for the impact parameter: $b\simeq\theta D_{OL}$, so that we can write the angular deflection as  
       \begin{equation}\label{defle}
       \alpha(\theta)=-\bar{a}\log\left(\frac{\theta D_{OL}}{b_c}-1\right)+\bar{b}\ .
       \end{equation}
       
       To obtain $\Delta\alpha_{n}$ we expand $\alpha(\theta)$ close to $\theta=\theta^{0}_n$, where $\alpha(\theta^{0}_n)=2n\pi$;
       \begin{equation}\label{da}
       \Delta\alpha_n=\frac{\partial\alpha}{\partial\theta}\Bigg|_{\theta=\theta^0_n}(\theta-\theta^0_n) \ .
       \end{equation}
       Evaluating (\ref{defle}) in $\theta=\theta^{0}_n$, gives
       \begin{equation}\label{To}
       \theta^0_{n}=\frac{b_c}{D_{OL}}\left(1+e_n\right), \qquad\text{where}\quad e_n=e^{\bar{b}-2n\pi} \ .
       \end{equation}
       Substituting (\ref{To}) in (\ref{da}), we get $\Delta\alpha_n=-\frac{\bar{a}D_{OL}}{b_ce_n}(\theta-\theta^0_n)$, and inserting that last result  into the lens equation (\ref{EqLente}), we get the expression for $n^{th}$ angular position of the image 
       \begin{equation}
       \theta_n\simeq\theta^0_n+\frac{b_ce_n}{\bar{a}}\frac{D_{OS}}{D_{OL}D_{LS}}(\beta-\theta^0_n) \ .
       \end{equation}
       Although the deflection of light preserves the surface brightness, the gravitational lens changes the appearance of the solid angle of the source. The total flux received from a gravitationally lensed image is proportional to magnification $\mu_{n}$, which is given by $	\mu_n=\left|\frac{\beta}{\theta}\frac{\partial\beta}{\partial\theta}|_{\theta^0_{n}}\right|^{-1}$. Then, using the (\ref{EqLente}) and recalling that $\Delta\alpha_n=-\frac{\bar{a}D_{OL}}{b_ce_n}(\theta-\theta^0_n)$, we get
       \begin{equation}
       \mu_{n}=\frac{e_n(1+e_n)}{\bar{a}\beta}\frac{D_{OS}}{D_{LS}}\left(\frac{b_c}{D_{OL}}\right)^2 \ .
       \end{equation}
        One must note that $\mu_n$ decreases very rapidly with $n$, so the brightness of the first image $\theta_1$ dominates in comparison with other ones. In any case, however, the luminosity is weak, due to the presence of the factor $\left(\frac{b_c}{D_{OL}}\right)^2$. We also observe that in the limit $\beta\to 0$ the magnification diverges, demonstrating that the perfect alignment of the source with the lens maximizes the possibility of detection of the relativistic images. Finally, we have expressed the position of the relativistic images as well as their fluxes in terms of the expansion coefficients ($\bar{a}$, $\bar{b}$, and $b_c$). Let us now consider the inverse problem, i.e., to determine the expansion coefficients from the observations. With this, we can understand the nature of the object generating the gravitational lens and compare the predictions made by modified gravity theories.   
       
       The impact parameter may be written in terms of $\theta_{\infty}$ \cite{Bozza2002},
       \begin{equation}
       	b_c=D_{OL}\theta_{\infty} \ .
       \end{equation}
       Let us follow Bozza \cite{Bozza2002} and assume that only the most external image $\theta_{1}$ is resolved as a single image while the others are encapsulated in $\theta_{\infty}$. Thus, Bozza defined the following observables,
       \begin{eqnarray}
       s&=&\theta_{1}-\theta_{\infty}= \theta_{\infty} e^{\frac{\bar{b}-2\pi}{\bar{a}}} \ ,\\
       \tilde{r}&=& \frac{\mu_{1}}{\sum\limits_{n=2}^{\infty} \mu_{n} }= e^{\frac{2\pi}{\bar{a}}} \ .
       \end{eqnarray}
      In the above expressions, $s$ is the angular separation, $\tilde{r}$ is the relationship between the flux of the first image and the flux of all the others. These forms can be inverted to obtain expansion coefficients. In the next subsection, we will give a concrete astrophysical example in order to calculate the observables aforementioned and measure the impact of the black-bounce parameter on them.
			
			\subsection{Application: black hole at the Galactic center}
			
			As said in section \ref{Int.}, observations data of stellar dynamics indicate the presence of a dark compact object at the center of our galaxy. This compact object is supposed to be a black hole and its mass is estimated being $4.4\times 10^{6}M_{\odot}$ \cite{Gezel2010}. We consider this astrophysical object in order to provide the behavior of the observables in terms of the dimensionless parameter $a/2M$.  
			
\begin{figure}
       	\subfloat[]{\includegraphics[width = 3in]{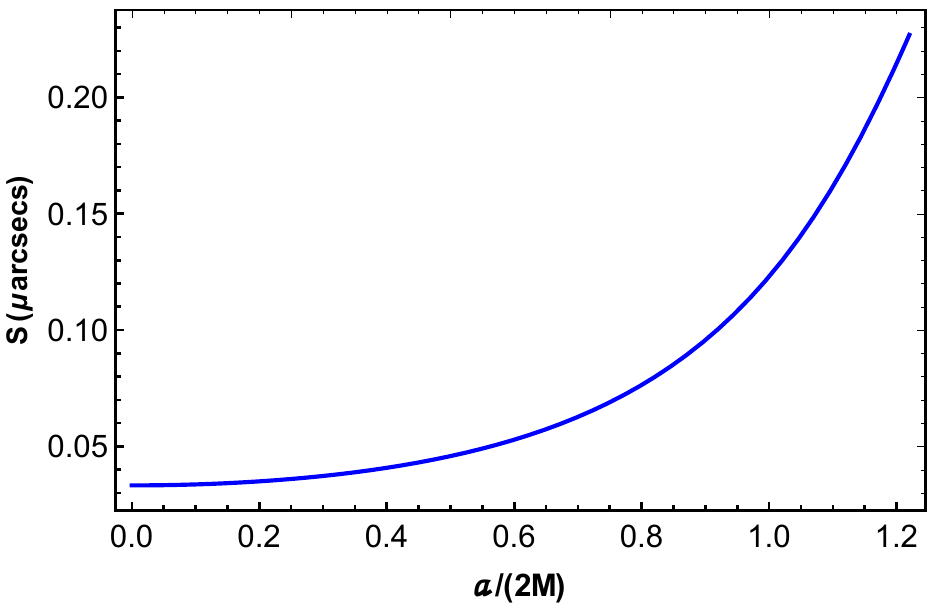}} 
       	\subfloat[]{\includegraphics[width = 3in]{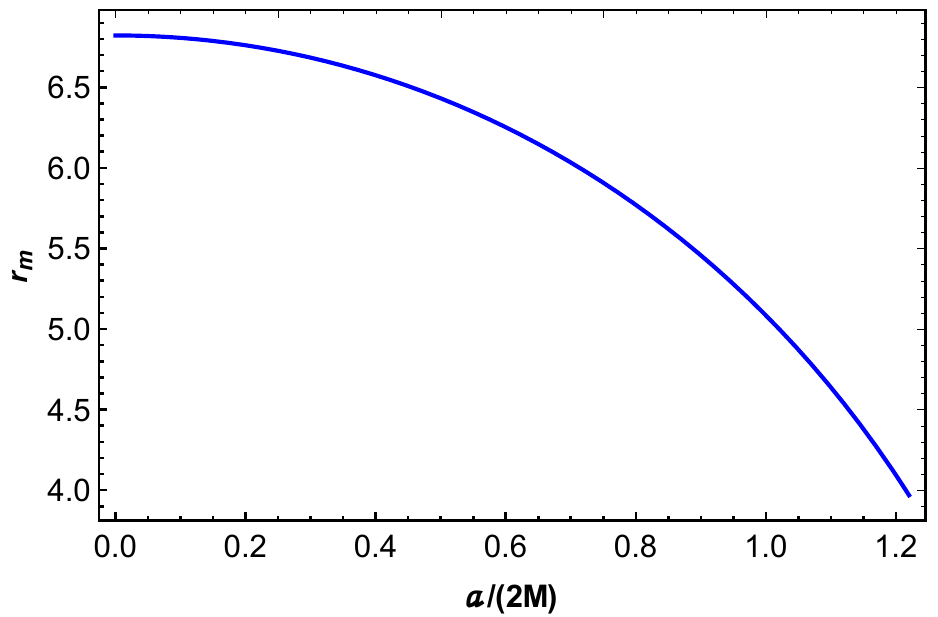}} 
       	\caption{Plot of the observables: (a), $s$, and (b), $r_m=2,5\log_{10}\tilde{r}$. }
       	\label{CE}
       \end{figure}

			To evaluate the observables, let us take distance $D_{OL}=8,5$Kpc, accordingly the literature. It should be noted that $b_c=3\sqrt{3}M$ does not depend on $a$ we can calculate it directly. From (\ref{bc}), we see that $\theta_{\infty}$ does not also depend on the black-bounce parameter, so we find $\theta_{\infty}\simeq26,5473 \ \mu\text{arcsecs}$. Conversely, the other parameters depend on $a$, explicitly. We plot the observables $s$ and $\tilde{r}$ (where we redefine it in terms of a logarithmic scale) in the Fig.\ref{CE} as functions of $\frac{a}{2M}$.
			As we can see, the angular separation increases with $a$, while $r_m$ decreases. This means that in relation to the Schwarzschild solution, the first image is further separated from the asymptotic images. In addition, the decrease in magnification with $a/2M$ implies that the first image is even less intense than the others. Note that, for the range $a/2M\geq 1$ the lensing parameters are associated with a wormhole. 

       \section{Conclusion}\label{conc}
       
        In this work, we calculated the angular deflection of light, in the weak and strong field limit, in the black-bounce spacetime. In the weak field limit, when the light passes too far from the photon sphere, we {showed that the lowest order in which the angular deflection of the light depends on the bounce parameter $a$ is quadratic, so for practical purposes, the angular deflection is equal to that of Schwarzschild spacetime. On the other hand, in the strong field limit, when the light passes very close to the photon sphere, the angular deflection differs significantly from the Schwarzschild case. We found its explicit form (\ref{angular}) in terms of the integral (\ref{IR}) which can only be calculated numerically. Next, we made a thorough numerical analysis of the angular deflection of light in terms of the impact factor $b$ for a set of different values of the bounce parameter $a$ and we concluded that the angular deflection grows as $a$ grows. In addition, our results recover the standard ones as expected, for example: Schwarzschild spacetime which means $a=0$. We also introduced angular deflection into the gravitational lens equation to evaluate the observables related to relativistic images. This was done by modeling our solution with the characteristics of the blackhole at the center of our galaxy \cite{Gezel2010}. Among the significant changes, we have shown that the angular separation between the first and the other relativistic images is greater than in the Schwarzschild case. In addition, we have shown that the first image is even less intense the others.

    \section*{Acknowledgments}
The work by A. Yu. P. has been supported by the CNPq project No. 301562/2019-9.

\end{document}